\documentclass[preprint,aps,prb,preprintnumbers]{revtex4}

\usepackage{amssymb}
\usepackage{amsmath}
\usepackage{dcolumn}
\usepackage{bm}
\usepackage{graphicx}
\usepackage{verbatim}
\usepackage{color}
\usepackage{textcomp}

\begin{document}

\title[Sample title]{Large-scale BN tunnel barriers for graphene spintronics}
\author{Wangyang Fu}
\author{P\'{e}ter Makk}
\author{Romain Maurand}
\author{Matthias Br\"{a}uninger}
\author{Christian Sch\"{o}nenberger}

\affiliation{Department of Physics, University of Basel, Basel,
Switzerland}

\date{\today}
\definecolor{darkgreen}{rgb}{.004,.68,.31}
\begin{abstract}

We have fabricated graphene spin-valve devices utilizing scalable
materials made from chemical vapor deposition (CVD). Both the
spin-transporting graphene and the tunnel barrier material are
CVD-grown. The tunnel barrier is realized by h-BN, used either as a
monolayer or bilayer and placed over the graphene. Spin transport
experiments were performed using ferromagnetic contacts deposited
onto the barrier. We find that spin injection is still greatly
suppressed in devices with a monolayer tunneling barrier due to
resistance mismatch. This is, however, not the case for devices with
bilayer barriers. For those devices, a spin relaxation time of
$\sim$260~ps intrinsic to the CVD graphene material is deduced. This
time scale is comparable to those reported for exfoliated graphene,
suggesting that this CVD approach is promising for spintronic
applications which require scalable materials.

\end{abstract}

\maketitle

% ----------------------------------------------------------------------------------------------------------------
 \section*{Introduction}
% ----------------------------------------------------------------------------------------------------------------

Owing to its small spin-orbit coupling, negligible hyperfine
interaction, and unparalleled high carrier mobility at room
temperature,~\cite{Novoselov04, Wang13, Seneor2012, Dlubak2012}
graphene is predicted to be the ideal material for spintronic
applications. Research conducted on graphene spintronics so far has
focused mostly on epitaxial or exfoliated high-mobility graphene
with various oxide-based tunnel barriers.~\cite{Dankert2014,
Birkner2013, Volk2013, Idzuchi2014, Tombros2007, Popinciuc2009,
Han2011, Yang2011, Droegeler2014, Zomer2012, Dlubak2012b,
Guimaraes2012} Spin relaxation times up to ns~\cite{Yang2011,
Droegeler2014} and spin transport in graphene over record distances
of 20~$\mu$m~\cite{Zomer2012} have been demonstrated in
four-terminal non-local measurement geometries at room temperature.
Very recently, spin relaxation lengths up to hundreds of $\mu$m have
also been reported in two-terminal devices (local measurement
geometry).~\cite{Dlubak2012, Seneor2012} However, in order to
achieve scalable devices, large-area graphene, as for example
provided by chemical vapor deposition (CVD) is
required.~\cite{Li2009} CVD graphene devices with Al$_{2}$O$_{3}$
tunnel barriers have been fabricated which exhibited spin relaxation
lengths comparable to those reported for exfoliated
graphene,~\cite{Avsar2011} showing their potential for future
applications.

Although these achievements are already quite remarkable, they are
still far from theoretical expectations based on the low spin-orbit
interaction and hyperfine coupling of graphene.~\cite{Guimaraes2012}
Several recent studies suggest that the performance of lateral
graphene spin valves is still limited by the quality of the tunnel
barriers in many devices.~\cite{Dlubak2012, Seneor2012, Idzuchi2014,
Droegeler2014, Sosenko2014} It is known that pin-holes can be formed
in oxide barriers due to inhomogeneous wetting of the oxide layer on
the hydrophobic graphene surface.~\cite{Dlubak2012b} This can
diminish the spin injection due to resistance mismatch. To achieve a
more homogeneous coverage, several groups have tried to
functionalize graphene before growing the oxide.~\cite{Dlubak2012b,
Wang2008, Robinson2010} However, this merely reduces the chances of
pin-hole formation but cannot rule them out. Pin-holes can still
appear. One of the solutions to this problem is to use a
two-dimensional (2D), atomically thin insulator with perfect
crystalline structure that can be placed onto
graphene.~\cite{Novoselov2005} Hexagonal boron nitride (h-BN), which
possesses a large band gap (5.97 eV) and has a small lattice
mismatch to graphene,~\cite{Wang13, Britnell2012, Meric2013} is a
promising candidate. Very recently, Yamaguchi et al. have
demonstrated exfoliated h-BN as tunnel barrier for spin injection
and detection in exfoliated lateral graphene spin valve
devices.~\cite{Yamaguchi2013}

In the present work, we combine layered tunnel barriers with
scalable fabrication by using large-area CVD h-BN as tunnel barriers
for CVD graphene-based lateral spin valves. We have performed both
local and non-local spin-valve measurements on the same devices, and
obtained spin relaxation times and tunnel barrier resistances
similar to those reported previously for exfoliated graphene samples
with exfoliated h-BN tunnel barriers, suggesting that CVD graphene
with CVD h-BN tunnel barrier is a promising system for large-scale
spintronic applications.

%----------------------------------------------------------------------------------------------------------------
\section*{Experimental methods}
% ----------------------------------------------------------------------------------------------------------------

To obtain graphene/h-BN heterostructures, we have performed a
sequential transfer of CVD graphene and CVD h-BN, followed by
electrode fabrication. We start with large-area CVD graphene, which
we have grown on Cu.~\cite{Fu2011} As shown in Fig.~1a, the graphene
layer was first transferred onto a Si substrate with a SiO$_2$ top
layer 285~nm in thickness and then patterned into micron-wide
ribbons using standard electron beam lithography (EBL) and oxygen
plasma etching. Then, a uniform monolayer (ML) of CVD
h-BN~\cite{GrapheneSupermarket} is transferred onto graphene using
the same transfer technique as for CVD graphene.~\cite{Li2009} Both
atomic force microscopy (AFM) and scanning electron microscopy (SEM)
images suggest the presence of a uniform ML h-BN with a multilayer
coverage of less than 20~$\%$. This is shown in Fig 1b, where the
red arrows indicate multilayer regions with triangular shape,
whereas the white ones show grain boundaries. A bilayer (BL) h-BN
tunnel barrier can be obtained by repeating the transfer process a
second time.

In order to minimize the oxidation of the ferromagnetic electrodes,
the Co strips (50~nm in thickness and 150 or 300~nm in width) are
placed in the final fabrication step. These strips are connected to
normal metal contacts made from Ti/Au (5~nm/45~nm) by an
intermediate thin Pd strip (10~nm). The non-magnetic contacts to
graphene are also made from similar thin Pd strips connected to
thicker Ti/Au leads. Fig.~1c shows a false-color SEM image of a CVD
graphene-based lateral spin valve device with a CVD h-BN tunnel
barrier. Four samples will be addressed in this study: a device with
an ML h-BN tunnel barrier and a graphene channel of length
$L=1$~$\mu$m denoted as ML1, and three BL h-BN devices with
different channel lengths of $L=$1, 2, and 3~$\mu$m, denoted as BL1,
BL2, and BL3. The three BL h-BN samples were fabricated together in
the same batch.

\begin{figure}
\includegraphics[width=140mm]{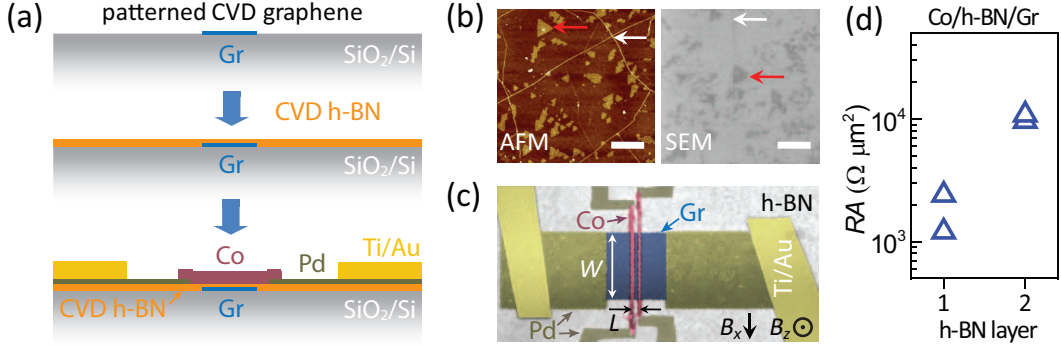}
  \caption{\label{Figure1}
(a) Process flow for the fabrication of CVD graphene-based lateral
spin valve devices with CVD h-BN tunnel barriers. (b) AFM and SEM
images of ML h-BN transferred onto a SiO$_{2}$/Si substrate. The
arrows highlight triangular-shaped ML h-BN (red) and h-BN grain
boundaries (white). Scale bar: 1~$\mu$m. (c) A false-color SEM image
of a fabricated device. The width $W$ of the graphene flake is
8~$\mu$m and the distance $L$ between the two ferromagnetic contacts
is 1~$\mu$m. The widths of the ferromagnetic Co electrodes are
150~nm and 300~nm. $B_{x}$ denotes the in-plane magnetic field
component parallel to the ferromagnetic contacts and $B_{z}$ the
out-of-plane one. (d) Dependence of the resistance-area product $RA$
on the number of transferred CVD h-BN layers in a Co/h-BN/graphene
junction.}
\end{figure}

% ----------------------------------------------------------------------------------------------------------------
\section*{Results and discussion}
% ----------------------------------------------------------------------------------------------------------------

We characterize the Co/h-BN/graphene junction resistances using
three-terminal measurements. Fig.~1d shows the obtained resistance-area
product $RA$ of
$\sim$1-3~k$\Omega$~$\mu$m$^{2}$ for sample ML1
(1~k$\Omega$ for contact areas of 2.4~$\mu$m$^{2}$ and 1.2~k$\Omega$ for contact areas of 1.2~$\mu$m$^{2}$)
and $\sim$10~k$\Omega$~$\mu$m$^{2}$ for sample BL1
(4~k$\Omega$ for 2.4~$\mu$m$^{2}$ and 9~k$\Omega$ for 1.2~$\mu$m$^{2}$), measured within a 1-10~mV bias
voltage range. These values are comparable to previously reported $RA$ products of exfoliated h-BN tunnel
barriers,~\cite{Yamaguchi2013} suggesting that we have successfully
fabricated tunnel barriers from CVD-grown h-BN between the
ferromagnetic layer and graphene.

At 4.2~K, all three BL h-BN samples show similar dependencies of the
resistivity of graphene $\rho_{G}$ on the back-gate voltage $V_{G}$
obtained by local four-terminal measurements. Fig.~2b depicts the
$\rho_{G}(V_{G})$ curve, as well as the corresponding conductivity
$\sigma(V_{G})=1/\rho(V_{G})$ (solid grey line), deduced from local
four-terminal measurement of the BL3 sample. At a back-gate voltage
$V_{G}$=0~V, we find $\rho_{G}$=3.6~k$\Omega$. To estimate the
carrier mobility $\mu$ we use the basic transport equation
$\sigma=en\mu$, where $e$ is the electron charge and $n$ the carrier
density. The charge density $en$ is proportional to the gate voltage
measured from the charge-neutrality point (CNP), $V_{CNP}$=14~V,
with a proportionality constant given by the gate capacitance $C$.
Explicitly, we obtain $\mu$ through the relation $g_{m}/{C}$, where
$g_m$ is the slope of the $\sigma(V_G)$ curve taken at negative gate
voltages far away from the CNP (see Fig.~2b solid red line). We
obtain $\mu\sim$~850~cm$^{2}$V$^{-1}$s$^{-1}$.

Fig.~2a shows the schematic circuit for the non-local
magneto-resistance (MR) measurements.~\cite{Johnson1985} A current
$I$ is injected from Co$_2$ to Pd$_2$ and a non-local voltage
$V_{nl}$ is measured between Co$_1$ and Pd$_1$. The non-local
resistance $R_{nl}$ is defined as the ratio of non-local voltage
$V_{nl}$ and the local current $I$, i.e. $R_{nl}=V_{nl}/I$. In this
setup, the charge current is separated from the voltage measurement
circuit, and only the spin current, which is injected at Co$_2$ and
diffusing outside of the current path, is detected by Co$_1$ at a
distance of $L$.

The in-plane magnetic field ($B_x$) dependence of the measured
non-local resistance $R_{nl}(B_x)$ for sample ML1 and BL1 at a gate
voltage of $V_{G}$=0~V is shown in Fig.~2c. Due to different widths
of the two ferromagnetic contacts (150~nm and 300~nm), the
ferromagnets switch at different $B_x$ fields, yielding a magnetic
field range in which the magnetic moments of the two contacts have
an antiparallel alignment. The difference in $R_{nl}$ observed for
parallel and antiparallel alignment defines the MR value. The solid
arrows indicate the polarization direction of Co$_1$ and Co$_2$, and
the horizontal arrows show the sweeping direction of the magnetic
field. Positive (negative) non-local voltage is observed when the
polarizations of the two ferromagnetic strips are parallel
(antiparallel).~\cite{vanSon1987} The measurements are performed
using the standard a.c. lock-in technique with a relatively large
supplied current $I_{DC}\sim$10~$\mu$A to obtain a good
signal-to-noise ratio. On sample ML1, we also performed non-local MR
measurements at 300~K. Clear non-local MR signals are observed at
both 4.2~K and 300~K (Fig.~2c), indicating electrical spin injection
from the ferromagnet to CVD graphene through the CVD h-BN tunnel
barriers over a wide temperature range. At 4.2~K, device ML1 with
low contact resistances ($\sim$1~k$\Omega$) exhibits a relatively
small non-local MR signal of only $\sim$40~m$\Omega$. With higher
h-BN thickness and otherwise identical parameters, device BL1
exhibits much higher contact resistances ($\sim$9~k$\Omega$) and
shows an in-plane non-local MR signal of $\sim$1.6~$\Omega$
(Fig.~2b). This suggests that the observed, relatively small
non-local MR signal in case of an ML CVD h-BN tunnel barrier is most
likely limited by resistance mismatch at the ferromagnetic contacts.

\begin{figure}
\includegraphics[width=170mm]{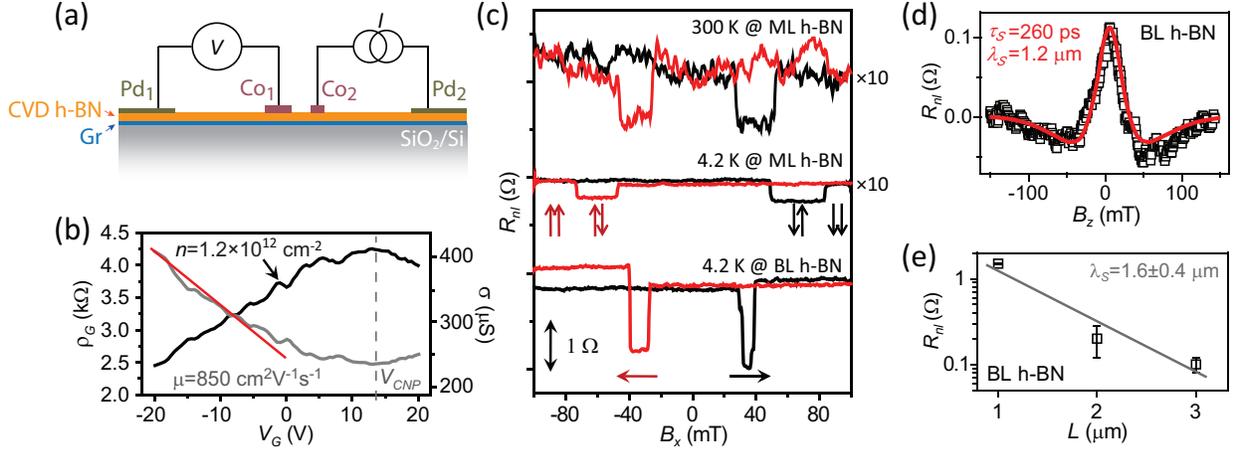}
  \caption{\label{Figure2}
(a) Schematic illustration of the device and the non-local MR
measurement configuration. (b) Resistivity $\rho_{G}$ of graphene as
a function of back-gate voltage $V_{G}$ (solid black curve) measured
at 4.2~K for the sample BL3 (BL h-BN with 3~$\mu$m graphene channel
length). The corresponding conductivity $\sigma(V_{G})$ is shown by
the solid grey curve. (c) In-plane non-local MR traces of the ML1
and BL1 samples (ML and BL h-BN with 1~$\mu$m graphene channel
length), measured at 300~K and 4.2~K with $I_{DC}\sim$10~$\mu$A at
$V_{G}$=0~V. The vertical arrows indicate the polarization direction
of Co$_1$ and Co$_2$, the horizontal arrows the sweep direction for
the magnetic field $B_{x}$. The non-local resistances of the ML1
sample (measured at 300~K and 4.2~K) have been multiplied by 10 for
clarity. (d) Hanle effect measured on device BL3 (graphene channel
of length 3 um) under similar conditions as in (c), using a
perpendicular magnetic field $B_{z}$. (e) The spin-dependent
non-local MR signal $\Delta$$R_{nl}$ on the Co electrode spacing $L$
at 4.2 K. The solid grey line represents the best fit based on
Eq.~(2).}
\end{figure}

A reliable determination of the spin relaxation time $\tau_{S}$ can
be accomplished by measuring the Hanle effect, the precession and
dephasing of spins in an out-of-plane magnetic field
$B_z$.~\cite{Johnson1985} We first polarized the ferromagnetic
electrodes in either parallel or antiparallel alignment by choosing
a suitable in-plane magnetic field $B_{x}$. Then, the non-local
resistances $R_{nl}$ were recorded as a function of $B_z$ for both
conditions. To eliminate any spurious contributions due to Hall
effect or other MR effects, we determined the Hanle curve by
subtracting the parallel from antiparallel measurement curves.
Fig.~2d depicts such a difference Hanle curve (at 4.2~K) for sample
BL3 measured at $V_{G}=$0~V. This curve can be fitted by using the
solution of the diffusion equation:~\cite{Jedema2002}

\[
\Delta R_{nl}\propto\int_{0}^{\infty}\frac{1}{\sqrt{4\pi
D_{S}t}}e^{-\frac{L^2}{4D_{S}t}}\cos\left(\frac{g\mu_{B}Bt}{\hbar}\right)e^{-\frac{t}{\tau_{S}}}dt
\eqno{(1)}
\]
where $D_{S}$ is the spin diffusion constant, $g$ the electron
g-factor set to 2 in the fit, $\mu_{B}$ the electron Bohr magneton,
$\hbar$ the reduced Planck constant, and $t$ the diffusion time. For
this fit, in addition to the g-factor, we fix $D_{S}$ by an estimate
obtained from the conductivity $\sigma$ using the Einstein relation
$\sigma = e^2 D_{S} \nu(E_F)$, where $\nu(E_F)$ is the graphene
density of states at the Fermi energy $E_F$. This later quantity can
be expressed by $n$ using the energy-momentum dispersion relation
$E(\vec{k})=\hbar v_F|\vec{k}|$ of idealized graphene with a Fermi
velocity $v_F$ of $10^6$~m/s. We obtain $D_{S}=$55~cm$^{2}$s$^{-1}$
at $V_{G}=$0~V where the carrier density is
$n=1.2\times10^{12}$~cm$^{-2}$.
The Hanle fit (solid red curve in Fig.~2d) results in a spin
relaxation time of $\tau_{S}=$260~ps, corresponding to a spin
relaxation length of $\lambda_{S}=\sqrt{D_{S}\tau_{S}}=$1.2~$\mu$m.
This is comparable to previous results obtained on CVD graphene as
well as exfoliated graphene samples on SiO$_{2}$/Si
substrate.~\cite{Popinciuc2009, Avsar2011}

One can also obtain an estimate for $\lambda_S$ by looking at the
absolute value of $\Delta R_{nl}$ for samples BL1-BL3 that were
fabricated together in one batch, but have different channel lengths
$L$. For these samples with bilayer h-BN tunneling barriers, the
contact resistances $R_c \sim 9$~k$\Omega$ are larger than the
graphene channel resistance determined over a spin-relaxation
length. This so called spin-relaxation resistance $R_{ch}^S$ is
given by $R_{ch}^{S}=\rho_{G}\lambda_{S}/W$, yielding $R_{ch}^S\sim
0.54$~k$\Omega$ with the width of the graphene flake $W=8$~$\mu$m.
In this regime, where resistance mismatch is not an issue, $\Delta
R_{nl}$ should follow the relation:~\cite{Jedema2002}
\[
\Delta
R_{nl}=P^{2}R_{ch}^{S}exp\left(-\frac{L}{\lambda_{S}}\right)
\eqno{(2)}
\]
where $P$ is the spin polarization of the injector contact. The measured three values for $\Delta R_{nl}$
are plotted in Fig.~2e on a logarithmic scale as a function of $L$. From the decaying
slope we deduce $\lambda_{S}=$1.6$\pm$0.4~$\mu$m at 4.2~K. This is in good agreement
with the relaxation length obtained from the non-local Hanle measurement discussed before.

\begin{figure}
\includegraphics[width=80mm]{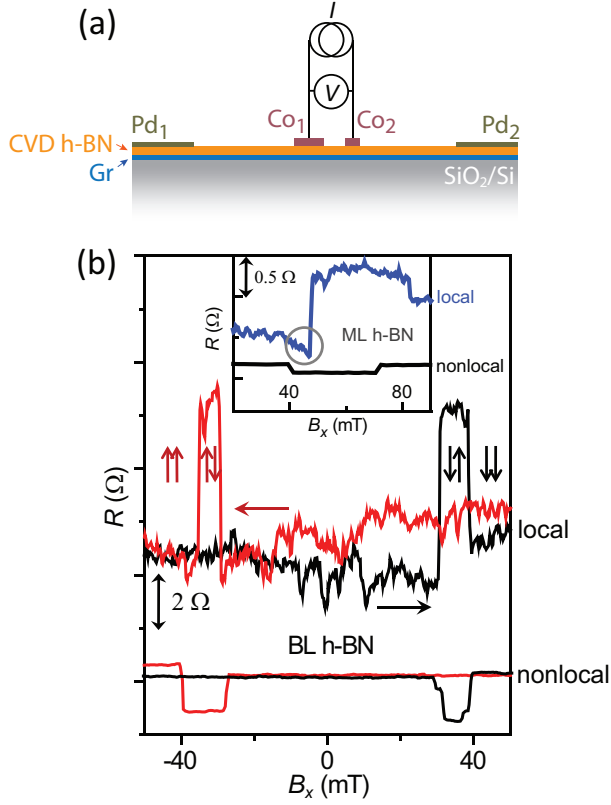}
  \caption{\label{Figure3}
(a) Schematic illustration of device and the local MR measurement
configuration. (b) In-plane local MR and non-local MR traces
measured at 4.2~K with $I_{DC}\sim$10~$\mu$A at $V_{G}$=0~V for
sample BL1 (BL h-BN with 1~$\mu$m graphene channel length). The
vertical arrows indicate the polarization directions of Co$_1$ and
Co$_2$, and the horizontal arrows show the sweep directions for the
magnetic field $B_{x}$. Inset, the MR of sample ML1 (ML h-BN with
1~$\mu$m graphene channel length) measured under similar conditions
but $V_{G}$=-32~V. The grey circle indicates possible AMR
contributions of the magnetic contacts to the local MR measurement.}
\end{figure}

After evaluating the non-local MR characteristics of the CVD
graphene-based lateral spin valve devices with CVD h-BN tunnel
barriers, we were also able to perform two terminal local MR
measurements on the same samples, see Fig. 3a.  Recently, Dlubak et
al.~\cite{Dlubak2012} deduced exceptionally long spin relaxation
lengths $\lambda_{S}$ exceeding hundreds of $\mu$m based on local MR
measurements on two terminal graphene spin valve devices with
graphene channels of only a few $\mu$m in length. However, they have
not performed non-local or Hanle measurements. It is therefore
instructive to compare non-local with local MR measurements. Fig.~3b
shows such a comparison on samples BL1 and ML1 (inset). The jumps in
the local resistances occurred at very similar switching fields to
those of the non-local resistance, but with opposite sign and
different magnitude. The difference in sign is what one expects in a
simplified model. For the local MR, an anti\-parallel alignment of
the magnetizations of the two Co strips will lead to a resistance
increase which is explained by the two channel current model,
similar to the tunneling MR (TMR) effect.~\cite{vanSon1987} In the
non-local measurement, a positive non-local voltage is expected if
both the injector and detector contact are aligned in parallel, and
a negative one in the opposite case. The local measurement also
exhibits a larger noise floor. While the spin and the charge current
paths are spatially separated in the non-local measurement, the spin
accumulation in the local geometry is detected in the presence of
the charge current background. This gives rise to the observed noise
floor as well as to an additional feature (grey circle in the inset
of Fig.~3b), which is possibly caused by the anisotropic MR (AMR)
contribution of the magnetic contacts.~\cite{Aurich2010}

For the interpretation of these measurements, we use the spin
relaxation length $\lambda_{S}=$1.2~$\mu$m deduced before, obtaining
$R_{ch}^{S}\sim$0.54~k$\Omega\ll$$R_C\sim$9~k$\Omega$ in the BL h-BN
devices, which suggests that spin relaxation inside the graphene
channel dominates. From the local MR measurements on sample BL1 we
found $\Delta R_{l,BL1}\sim$6.6~$\Omega$ at $V_{G}$=0~V. This value
is about 4 times as large as the non-local MR spin signal of $\Delta
R_{nl,BL1}\sim$1.6~$\Omega$ at the same gate voltage. This is
interesting as the simple theory would predict only a factor of 2
difference.~\cite{Seneor2012} The measurements on sample ML1 shown
in the inset of Fig.~3b yield a similar result: $\Delta
R_{l,ML1}\sim$0.4~$\Omega$ and $\Delta R_{nl,ML1}\sim$0.1~$\Omega$,
which also amounts to a factor of 4. The explanation of this result
requires further studies.

% ----------------------------------------------------------------------------------------------------------------
\section*{Conclusions}
% ----------------------------------------------------------------------------------------------------------------

In conclusion, we have assembled scalable spin valve devices using
large-scale CVD graphene as transport material and large-scale CVD
h-BN as tunnel barrier. During our studies of spin-transport
properties, we have found spin relaxation times comparable to those
reported for exfoliated graphene samples fabricated on SiO$_{2}$/Si
substrates. Moreover, we have shown that both local and non-local
measurements can be performed on the same device. Further
investigations in this direction can shed light on the spin
relaxation mechanism which is still not completely understood in
graphene. We believe that this work paves the way for further
research on graphene spintronic devices based on large-scale 2D
heterostructures. After finishing the manuscript we became aware of
a similar work on exfoliated graphene with CVD h-BN tunnel
barriers.~\cite{VenkataKamalakar2014}

\begin{acknowledgments}
The authors acknowledge funding from the Swiss National Science
Foundation (SNF), ESF Euro\-graphene, NCCR-QSIT, the FP7 project
SE2ND, ERC QUEST, the Swiss Nano\-science Institute (SNI) and
Graphene Flagship. The authors also wish to thank Bernd Beschoten,
Takis Kontos, Andreas Baumgartner, J\"{o}rg Gramich, Julia Samm, and
Simon Zihlmann for helpful discussions. Wangyang Fu and P\'{e}ter
Makk contributed equally to this work.
\end{acknowledgments}

% ----------------------------------------------------------------------------------------------------------------
% REFERENCES
% ----------------------------------------------------------------------------------------------------------------

\end{document}